\documentclass[twoside,twocolumn,9pt]{article}
\usepackage{extsizes}
\usepackage[super,sort&compress,comma]{natbib} 
\usepackage[version=3]{mhchem}
\usepackage[left=1.5cm, right=1.5cm, top=1.785cm, bottom=2.0cm]{geometry}
\usepackage{balance}
\usepackage{mathptmx}
\usepackage{sectsty}
\usepackage{graphicx} 
\usepackage{lastpage}
\usepackage{bm}
\usepackage{color}
\usepackage[format=plain,justification=justified,singlelinecheck=false,font={stretch=1.125,small,sf},labelfont=bf,labelsep=space]{caption}
\usepackage{float}
\usepackage{fancyhdr}
\usepackage{fnpos}
\usepackage[english]{babel}
\addto{\captionsenglish}{%
  
}
\usepackage{array}
\usepackage{droidsans}
\usepackage{charter}
\usepackage[T1]{fontenc}
\usepackage[usenames,dvipsnames]{xcolor}
\usepackage{setspace}
\usepackage[compact]{titlesec}
\usepackage{hyperref}

\usepackage{epstopdf}
\definecolor{cream}{RGB}{222,217,201}
\usepackage{color}

\begin{document}

\pagestyle{fancy}
\thispagestyle{plain}
\fancypagestyle{plain}{
\renewcommand{\headrulewidth}{0pt}
}

\makeFNbottom
\makeatletter
\renewcommand\LARGE{\@setfontsize\LARGE{15pt}{17}}
\renewcommand\Large{\@setfontsize\Large{12pt}{14}}
\renewcommand\large{\@setfontsize\large{10pt}{12}}
\renewcommand\footnotesize{\@setfontsize\footnotesize{7pt}{10}}
\makeatother

\renewcommand{\thefootnote}{\fnsymbol{footnote}}
\renewcommand\footnoterule{\vspace*{1pt}%
\color{cream}\hrule width 3.5in height 0.4pt \color{black}\vspace*{5pt}} 
\setcounter{secnumdepth}{5}

\makeatletter 
\renewcommand\@biblabel[1]{#1}            
\renewcommand\@makefntext[1]%
{\noindent\makebox[0pt][r]{\@thefnmark\,}#1}
\makeatother 
\renewcommand{\figurename}{\small{Fig.}~}
\sectionfont{\sffamily\Large}
\subsectionfont{\normalsize}
\subsubsectionfont{\bf}
\setstretch{1.125} 
\setlength{\skip\footins}{0.8cm}
\setlength{\footnotesep}{0.25cm}
\setlength{\jot}{10pt}
\titlespacing*{\section}{0pt}{4pt}{4pt}
\titlespacing*{\subsection}{0pt}{15pt}{1pt}

\fancyfoot{}
\fancyfoot[LO,RE]{\vspace{-7.1pt}\includegraphics[height=9pt]{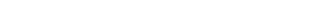}}
\fancyfoot[CO]{\vspace{-7.1pt}\hspace{13.2cm}\includegraphics{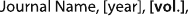}}
\fancyfoot[CE]{\vspace{-7.2pt}\hspace{-14.2cm}\includegraphics{head_foot/RF}}
\fancyfoot[RO]{\footnotesize{\sffamily{1--\pageref{LastPage} ~\textbar  \hspace{2pt}\thepage}}}
\fancyfoot[LE]{\footnotesize{\sffamily{\thepage~\textbar\hspace{3.45cm} 1--\pageref{LastPage}}}}
\fancyhead{}
\renewcommand{\headrulewidth}{0pt} 
\renewcommand{\footrulewidth}{0pt}
\setlength{\arrayrulewidth}{1pt}
\setlength{\columnsep}{6.5mm}
\setlength\bibsep{1pt}

\makeatletter 
\newlength{\figrulesep} 
\setlength{\figrulesep}{0.5\textfloatsep} 

\newcommand{\topfigrule}{\vspace*{-1pt}%
\noindent{\color{cream}\rule[-\figrulesep]{\columnwidth}{1.5pt}} }

\newcommand{\botfigrule}{\vspace*{-2pt}%
\noindent{\color{cream}\rule[\figrulesep]{\columnwidth}{1.5pt}} }

\newcommand{\dblfigrule}{\vspace*{-1pt}%
\noindent{\color{cream}\rule[-\figrulesep]{\textwidth}{1.5pt}} }

\makeatother

\twocolumn[
  \begin{@twocolumnfalse}
{\includegraphics[height=30pt]{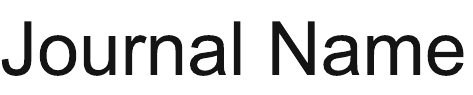}\hfill\raisebox{0pt}[0pt][0pt]{\includegraphics[height=55pt]{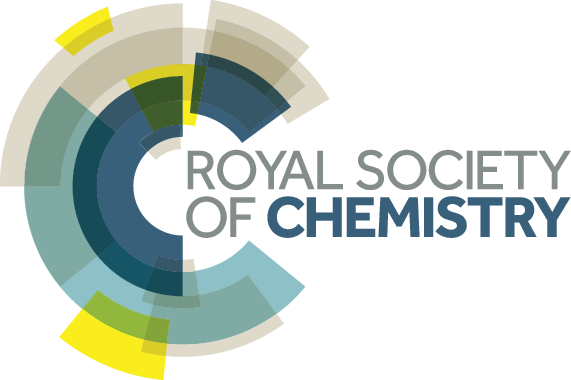}}\\[1ex]
\includegraphics[width=18.5cm]{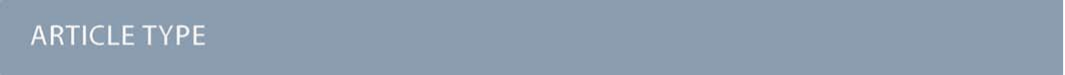}}\par
\vspace{1em}
\sffamily
\begin{tabular}{m{4.5cm} p{13.5cm} }

\includegraphics{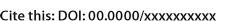} & \noindent\LARGE{\textbf{Curvature induces and enhances transport of spinning colloids through narrow channels$^\dag$}} \\
\vspace{0.3cm} & \vspace{0.3cm} \\
 & \noindent\large{
 Eric Cereceda-L\'opez,\textit{$^{a, b}$} 
 Marco De Corato,\textit{$^{c}$} 
 Ignacio Pagonabarraga,\textit{$^{a, d}$} 
Fanlong Meng, \textit{$^{e,f,g}$}
 Pietro Tierno,\textit{$^{*,a, b, d}$} and 
 Antonio Ortiz-Ambriz,\textit{$^{*,h}$}
 } \\
\includegraphics{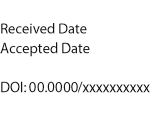} & \noindent\normalsize{The effect of curvature and how it induces and enhances the transport of colloidal particles driven through narrow channels represent an unexplored research avenue. Here we combine experiments and simulations to investigate the dynamics of 
magnetically driven colloidal particles confined through a narrow, circular channel. We use an external precessing magnetic field to induce a net 
torque and spin the particles at a defined angular velocity. Due to the spinning, the particle propulsion emerges from the different hydrodynamic coupling with the inner and outer walls and strongly depends on the curvature. The experimental findings are combined with finite element numerical simulations that predict a positive rotation translation coupling in the mobility matrix. Further, we explore the collective transport of many particles across the curved geometry, making an experimental realization of a driven single file system.  With our finding, we elucidate the effect of curvature on the transport of microscopic particles which could be important to understand the complex, yet rich, 
dynamics of particle systems driven through curved microfluidic channels. } \\
\end{tabular}

 \end{@twocolumnfalse} \vspace{0.6cm}
 ]

\renewcommand*\rmdefault{bch}\normalfont\upshape
\rmfamily
\section*{}
\vspace{-1cm}


\footnotetext{\textit{
$^{a}$~Departament de F\'{i}sica de la Mat\`{e}ria Condensada, Universitat de Barcelona, 08028, Spain.
E-mail: ptierno@ub.edu}}
\footnotetext{\textit{
$^{b}$~Institut de Nanoci\`{e}ncia i Nanotecnologia, Universitat de Barcelona (IN2UB), 08028, Barcelona, Spain.}}
\footnotetext{\textit{
$^{c}$~Aragon Institute of Engineering Research (I3A), University of Zaragoza, Zaragoza, Spain.}}
\footnotetext{\textit{
$^{d}$~University of Barcelona Institute of Complex Systems (UBICS), 08028, Barcelona, Spain}}
\footnotetext{\textit{
$^{e}$~Chinese Academy of Sciences, CAS Key Laboratory for Theoretical Physics, Institute of Theoretical Physics, Beijing 100190, China
}}
\footnotetext{\textit{
$^{f}$~University of Chinese Academy of Sciences, School of Physical Sciences, Beijing 100049, China
}}
\footnotetext{\textit{
$^{g}$~University of Chinese Academy of Sciences, Wenzhou Institute, Wenzhou 325000, China 
}}
\footnotetext{\textit{
$^{h}$~Tecnologico de Monterrey, School of Science and Engineering, 64849, Monterrey, Mexico. E-mail: aortiza@tec.mx}}
\footnotetext{\dag~Electronic Supplementary Information (ESI) available: Two videos illustrating the single and collective 
particle dynamics within the ring. See DOI: 00.0000/00000000.}



\section{Introduction}

In viscous fluids the Reynold ($Re$) number (the ratio between inertial and viscous forces) is relatively low and the Navier-Stokes equations become effectively time-reversible.~\cite{Brenner} Under such conditions, 
inducing the transport of microscale objects may become problematic since any reciprocal motion, namely a 
series of body/shape deformations that are identical when reversed,
do not produce a net movement.~\cite{purcell_life_1977}
However, many artificial prototypes are often powered by external fields, such as magnetic ones, which indeed induce periodic deformation or translations. Thus, at low $Re$ these systems require a subtle strategy to overcome the time-reversibility and propel, such as the presence of flexibility,~\cite{Dreyfus2005,Gao2010,Cebers2016,Yang2020} non linearity of the dispersing medium,~\cite{Quiu2014,Li2021}  
or the interaction between different parts.~\cite{Snezhko2016,Torres2018,Steinbach2020} 

The problem of propulsion at low $Re$ becomes even more complex when the active object, either a microswimmer or a synthetic self-propelled particle, is confined within a narrow channel.~\cite{bechinger_active_2016} The presence of confining walls can affect the microswimmer dynamics in many ways. In particular, planar surface can attract swimming bacteria 
due to hydrodynamic interactions,~\cite{frymier_threedimensional_1995, berke_hydrodynamic_2008}
induce circular trajectories,~\cite{lauga_swimming_2006,Leonardo2011}
or curved posts can be used as reflecting barriers or traps.~\cite{spagnolie_geometric_2015} 
Moreover, it has been recently found that the curvature of the fallopian tubes guides the locomotion of sperm cells to promote fertilization,~\cite{raveshi_curvature_2021} and similar situations are not limited to bacteria or sperm cells, but extend to other types of situations.~\cite{bianchi_brownian_2020, kuron_hydrodynamic_2019,volpe_microswimmers_2011,spagnolie_hydrodynamics_2012}

On the other hand, confinement can be used to rectify the random motion of microswimmers using patterned channels and other ratcheting methods.~\cite{Galajda2007,Wan2008,ai_transport_2014, yariv_ratcheting_2014, koumakis_directed_2014, pototsky_rectification_2013, ai_rectification_2013, guidobaldi_geometrical_2014, kantsler_ciliary_2013} 
At low $Re$, the close proximity of a flat wall can break the spatial symmetry of the flow field and induce propulsion in simpler artificial systems, such as rotating 
magnetic particles,~\cite{tierno_controlled_2008,Driscoll2017,martinez-pedrero_emergent_2018,Tierno2021} doublets \cite{Morimoto2008} or elongated structures.~\cite{Sing2010,Zhang2010}

Here we show that the geometry of the channel can be used to induce translation of a driven particle by breaking the rotational symmetry of the hydrodynamic interactions. We confine spinning magnetic colloids
within narrow circular channels characterized by a width close to the particle diameter. We find that curvature can rectify the particle rotational motion inducing a net translational one. Moreover, these surface rotors are able to 
translate along the ring at a velocity that depends on the curvature of the channel. We explain this experimental observation using numerical simulations which elucidate the mechanism of motion based on the difference between the hydrodynamic coupling with the two confining walls, where the largest coupling is with the wall characterized by the largest curvature. 
Further, we investigate the collective 
propulsion of many driven rotors, and show  
that within the circular ring the magnetic particles behave as 
a driven single-file, with an initial diffusive regime 
followed by a ballistic one at a large time scale.

\begin{figure}[t]
\centering
\includegraphics[width=\columnwidth,keepaspectratio]{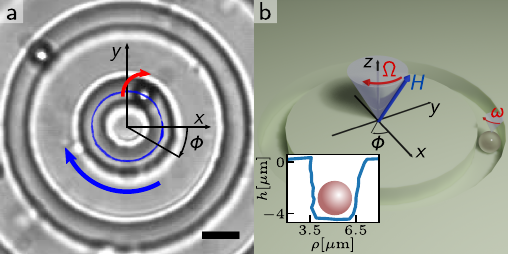}
\caption{(a) Optical microscope image of two circular channels  
in a PDMS mold each with a spinning paramagnetic colloidal particle. The red (blue) arrow in the image indicates
the direction of the particle rotation (trajectory). Scale bar is $5 \rm{\mu m}$, see also corresponding 
Movie S1\dag  in the Supporting Information.
(b) Schematic of a particle within the 
circular channel and of the applied precessing magnetic field $\bm{H}$. The inset shows the transverse profile of the channel measured using confocal profilometry.}
\label{fig:figure1}
\end{figure}
\section{Methods}

\subsection{Experimental protocol}

The circular microchannels are realized in Polydimethylsiloxane (PDMS) with soft lithography. We first use commercial software (AUTOCAD) to design a sequence of lithographic channels with a width  $w=  3 \, \rm{\mu m} $,
see Fig.~\ref{fig:figure1}(a). The channels are placed concentrically so that they are always separated by a distance of $10\rm{\mu m}$. To sample the channels in 5$\rm{\mu m}$ intervals,  the rings are separated in two sets, one starting at $5 \, \rm{\mu m}$ and the other at $10 \, \rm{\mu m}$. 
The microchannels present a depth of $\sim 4 \rm{\mu m}$ thus, larger than the particle width, such that the particles are completely surrounded by solid walls, as shown by the confocal profile in Fig.~\ref{fig:figure1}(b).

We use the following procedure to realize the structures in PDMS. First, we fabricate reliefs in SU-8, 
an epoxy-based negative photoresist, on top of a silicon (Si) wafer. The latter was prepared by dehydrating it for $10$ minutes at $200^{\circ}$ C on a hot plate and then plasma cleaned it for $1$ minute at a pressure of $1$ Torr. After that, a layer of SU-8 $3005$ is spin-coated on top of the Si wafer for $10$ s at $500$ revolution per minute (rpm), followed by spinning at $4000$ rpm for 
$120$ s, and later baked on a hot plate heated at a temperature of $95^{\circ}$ C for $2$ minutes. 
The SU-8 is then polymerized by exposure to 
UV light through a Cromium (Cr) mask for $5.6$ s with a light intensity of $100 \, \rm{mJ \, cm^{-2}}$ (Mask aligner, i-line configuration with wavelength $\lambda$=365nm).
Then the exposed film is baked again for $1$ min at $65^{\circ}$ C and for another $1$ min at $95^{\circ}$ C, and developed for $30$ s in propylene glycol methyl ether acetate.
We then transfer the obtained master to PDMS
by first cleaning the SU-8 structure in a plasma cleaner, operating at a pressure of $1$ Torr for $1$ min, and then silanizing it in a vacuum chamber with a drop of silane (SiH$_4$) for one hour. After that, the PDMS is spin coated onto the Si wafer at $4000$ rpm during $1$ min and baked for $30$ min above a hot plate at a temperature of $95^{\circ}$ C. The PDMS is then transferred to a cover glass by plasma cleaning both surfaces ($1$ min at $1$ Torr), then joining the PDMS and the glass and finally peeling very carefully with a sharp blade the resulting PDMS film. 

Once the PDMS channels are fabricated, we realize a closed fluidic cell
by first painting the PDMS patterned coverglass
with two strips of silicon-based vacuum grease. After that, we
deposit a drop containing the colloidal particles dispersed in water
on top of it, we close it with another clean coverglass and then we seal the two open sides with more grease. The colloidal suspension is prepared by dispersing $1.5 \, \rm{\mu l}$ from the stock solution of 
paramagnetic colloidal particles (Dynabeads M-270, $2.8\rm{\mu m}$ in diameter) in $1 \, \rm{ml}$ of ultrapure water (MilliQ system, Millipore). 

The fluidic cell is placed on a custom-made inverted optical microscope equipped with optical tweezers that are controlled using two Acousto-Optic Deflectors (AA Optoelectronics DTSXY-400-1064) driven by a radio frequency wave generator (DDSPA2X-D431b-34). 
The optical tweezers ensure that the channels are filled with the desired number of particles. Since it is often necessary to push the deposited particles out of the channels, the tweezers have been set up to trap from below by driving the laser radiation (Manlight ML5-CW-P/TKS-OTS) through a high numerical aperture objective (Nikon $100\times$, Numerical aperture $1.3$). The trapping objective is also used for observation and a custom-made TV lens is mounted in the microscope setup to observe a wide enough field of view, $\sim 100\time 100 \, \rm{\mu m^2} $. 

The external magnetic fields used to spin the particles are applied using 
a set of five coils, all arranged around the sample. In particular,  two coils have the main axis aligned along the $\hat{\bm{x}}$ direction, two for the $\hat{\bm{y}}$ direction and one for the $\hat{\bm{z}}$ direction which is perpendicular to the particle plane ($\hat{\bm{x}},\hat{\bm{y}}$). The sample is placed directly above the coil oriented along the $\hat{\bm{z}}$ axis. 
The magnetic coils are connected to three power amplifiers (KEPCO BOP 20-10) which are driven by a digital-analog card (cDAQ NI-9269) and controlled with a custom-made LabVIEW program.

\subsection{Numerical simulations}

To understand the mechanism of propulsion of the rotating particles in the channel, we perform finite element simulations. We consider a sphere suspended in a Newtonian liquid with viscosity $\eta$ and confined in a channel obtained from two concentric cylinders of height $h$, as schematically depicted in Fig. \ref{fig:figure3} (a). The Reynolds number, estimated from the characteristic particle velocity in the ring, is vanishingly small and thus we neglect inertial effects. We consider a Cartesian reference frame with origin at the center of the channel (Fig.~\ref{fig:figure1}). We assume that the distance between the particle and the bottom wall, $z_p$, is fixed due to the balance between gravitational forces and electrostatic repulsion. The ratio of the lateral size of the channel $w$ and the particle radius is fixed to ${w}/{a}=2.2$ which for our $a=1.4\rm{\mu m}$ particles produces a channel $3.1\rm{\mu m}$ wide, similar to that shown in the inset of \ref{fig:figure1}b). To model an open channel in the experimental system, we chose $h$ much larger than $w$. The particle is placed near the bottom of the domain, and horizontally at the center, as depicted in Fig. \ref{fig:figure3} (a). The rotating magnetic field generates a constant torque $\bm{\tau}_m$ along the $\hat{\bm{z}}$ axis, which leads to a constant angular velocity $\omega_p$ along the same direction. 
As a result of the rotation, we expect the particle to translate along the direction that is tangential to the channel centerline, $\hat{\bm{\phi}}$, due to the hydrodynamic coupling with the channel walls (see Fig. \ref{fig:figure1} ).

To investigate how the motion of the particle along the channel depends on the radius of curvature, we compute the mobility matrix of the particle. The mobility matrix, $\bm{M}$, relates the translational ($\bm{v}_p$) and angular velocity ($\bm{\omega}_p$) of the particle to external forces ($\bm{F}$) and torques ($\bm{\tau}$): 
\begin{align}
\begin{bmatrix}
	\bm{v}_p \\ \bm{\omega}_p
\end{bmatrix} = 
\bm{M} 
\begin{bmatrix}
	\bm{F} \\ \bm{\tau}
\end{bmatrix}   \,  \, \, \, .
\end{align}
The mobility matrix, $\bm{M}$, contains all the information about the hydrodynamic interactions between the particle and the surrounding boundaries. In this particular case, $\bm{M} $ depends on the radius of the channel $R$. Here, we are interested in the coefficient that couples the magnetic torque, acting along the $\hat{\bm{z}}$ direction, to the translational velocity along the $\hat{\bm{\phi}}$ direction. 

To obtain $\bm{M}$, we first find the resistance matrix  $\bm{R}$, and then compute $\bm{M}$ as  $\bm{M}=\bm{R}^{-1}$. To do so, we solve the Stokes equation by imposing a single component of the velocity vector and then computing the resulting force and torque acting on the particle:
\begin{align}
\nabla^2\bm{v}-\nabla p = 0\\
\nabla\cdot\bm{v} = 0
\end{align}
where $\bm{v}$ is the velocity of the fluid and $\bm{p}$ is the pressure. 
We consider no-slip boundary condition on the walls of the domain $\bm{v}|_\mathcal{S}=0$, being
$\mathcal{S}$ the surface of the particle.
On $\mathcal{S}$ we fix the velocity or the angular velocity. The total forces and torques acting on the particle are computed using:
\begin{align}
\int_\mathcal{S} \bm{T}\cdot\bm{n}\  \mathrm{d}S = \bm{F}\\
\int_\mathcal{S} \left(\bm{r}-\bm{r_p}\right)\times\left(\bm{T}\cdot\bm{n}\right)\  \mathrm{d}S = \bm{\tau}
\end{align}
where $\bm{r}_p = (x_p, y_p, z_p)$ is the position vector of the particle center, $\bm{r}$ is the position of a point on the surface of the particle and $\bm{T}$ is the stress tensor defined as: $\bm{T} = \eta(\nabla v+\nabla v^T) - p\bm{I}$, with $p$ the pressure.
By repeating the simulation for all the six components of the velocity vector, we can construct the resistance matrix $\bm{R}$ which is the inverse of the mobility matrix, $\bm{M} = \bm{R}^{-1}$.

\begin{figure}[t]
\centering
\includegraphics[width=\columnwidth,keepaspectratio]{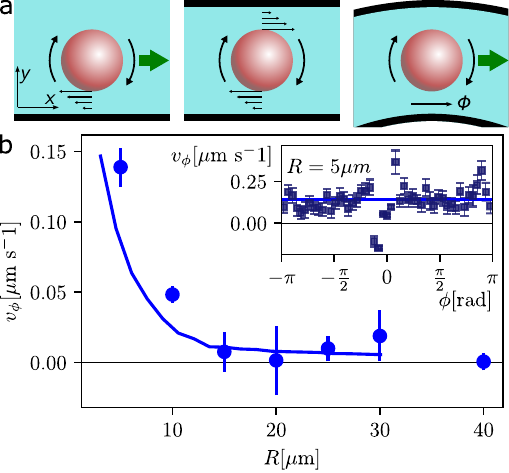}
\caption{(a) Schematic showing a spinning particle close to a single, flat  wall (left), confined between 
two flat walls (center) and between two curved walls (right). All three images show the particles from the top view, i.e. the particle rotates in the ($\hat{\bm{x}},\hat{\bm{y}}$) plane. The bottom wall, placed at $z=0$, is not shown for clarity. (b) Tangential velocity $v_{\phi}$ of the rotating magnetic particle inside the channels for different values of the central radius $R$. The filled points are experimental measurements, and the continuous line is calculated from the numerical simulations using a magnetic torque equal to $\tau_m = 0.55 \, \rm{pN \, \mu m}$. The inset shows the mean velocity at each value of the angle for a ring with radius $R=5 \, \rm{\mu m}$. The angle $\phi$ is measured with respect to the $\hat{\bm{x}}$ axis.}
\label{fig:figure2}
\end{figure}

\begin{figure*}[th]
\centering
\includegraphics{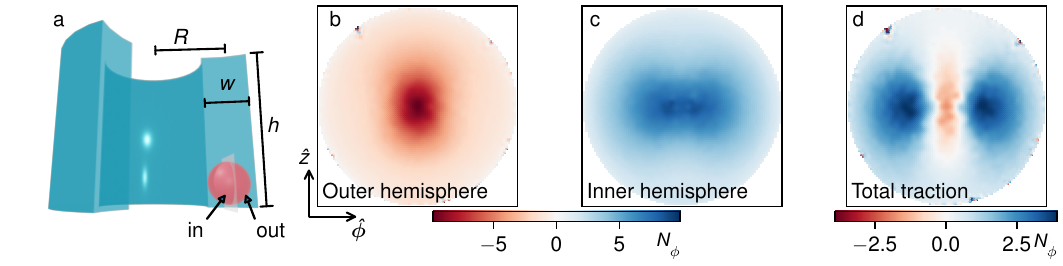}
\caption{(a) Schematic showing the simulation domain considered, which consists of the space between two concentric cylinders separated by a spacing $w$, with height $h$, and a central radius $R$. (b,c) Surface traction in the outer (b) and inner (c) hemispheres of the particle, as defined in panel (a). (d) Total traction given by the sum of the functions shown in panels (b) and (c), which is positive, indicating a resulting force that pushes forward the particle. }
\label{fig:figure3}
\end{figure*}

\section{Single particle motion}
We start by illustrating the dynamics of individual spinning particles dispersed within circular channels characterized by different radius $R$. 
Fig.~\ref{fig:figure1}(a) shows a microscope image of two  channels with radii $R= 5,15 \, \rm{\mu m}$ and filled with one paramagnetic colloidal particle each,  
while the corresponding schematic of the experimental system is shown 
in Fig.~\ref{fig:figure1}(b).  
The spinning is induced by the application of an external magnetic field 
that performs conical precession at a frequency $\omega$ and
an angle $\theta$ around the $\hat{\bm{z}}$ axis:
\begin{equation}
\bm{H} \equiv 
H_0 [\cos{\theta} \hat{\bm{z}} +
\sin{\theta}(\cos{(\omega t)} \hat{\bm{x}} + 
\sin{(\omega t)} \hat{\bm{y}})
]  \, \, \, ,
\label{field}
\end{equation}
being $H_0$ the field amplitude.
The precessing field has an in-plane component 
$\bm{H_{xy}}\equiv \sin{\theta}(\cos{(\omega t)} \hat{\bm{x}} + 
\sin{(\omega t)} \hat{\bm{y}})$
which is used to spin the particles, and a static one perpendicular to the particle plane, $\bm{H}_z=H_0 \cos{\theta} \hat{\bm{z}}$
which is used to control the dipolar interactions when many particles are present. In particular, we keep constant the magnetic field values to $H_0=900 \, \rm{A\, m^{-1}}$, the frequency $f=50$Hz
such that $\omega=2\pi f=314.2 \, \rm{rad \, s^{-1}}$
and the cone angle at $\theta=44.5^{\circ}$. 

For the single particle case, 
the precessing field at relative high frequency, is 
able to induce 
an average magnetic torque, $\bm{\tau}_m=\langle \bm{m} \times \bm{H} \rangle$,
which sets the particles
in rotational motion at an angular speed $\omega_p$ close to the plane. 
This torque arises due to the finite relaxation time $t_r$ of the particle magnetization, 
as demonstrated in previous works.~\cite{Tierno2007,Janssen2009} 
Note that here, for simplicity, we assume that the particles are characterized by a single relaxation time.
For a paramagnetic colloid with radius $a=1.4 \, \rm{\mu m}$ and magnetic volume susceptibility $\chi\sim 0.4$, the torque can be calculated as~\cite{Cebers2011,martinez-pedrero_colloidal_2015}, 
\begin{equation}
\bm{\tau}_m=\frac{4a^3\pi \mu_0  \chi H_0^2\sin^2{(\theta)}\omega t_r}{3(1+t_r^2\omega^2)}\hat{\bm{z}}
\label{torque}
\end{equation}
being $\mu_0=4 \pi 10^{-7} \, \rm{H \, m}$ the permeability of vacuum, similar to that of water. 
Assuming $t_r\sim 10^{-3}$ s we obtain $\tau_m= 0.32 \, \rm{pN \, \mu m}$.

In the overdamped limit, the magnetic torque is balanced by the viscous  torque $\bm{\tau}_v$ arising from
the rotation in water,  $\bm{\tau}_m+\bm{\tau}_v=\bm{0}$. Close to a planar wall, 
$\bm{\tau}_v$ can be written as, $\bm{\tau}_v=-8\pi \eta \bm{\omega}_p a^3 f_r(a/z_p)$, where $\eta=10^{-3} \, \rm{Pa \cdot s}$ is the viscosity of water, and $f_r(a/z_p)$ is a correction term that accounts for the proximity of the bottom wall at a distance $z_p$.~\cite{Goldman1967} 
In particular, assuming the space between the particle and the wall $\sim 100$ nm, $f_r(a/h)=1.45$. In the overdamped regime, the balance between both torques allows estimating the particle mean angular velocity: $\langle \omega_p \rangle = \mu_0 H_0^2\sin^2{\theta} \chi \tau_r \omega/[6\eta (1+\tau_r^2 \omega^2)]$, which gives $\omega_p=3.27 \rm{rad \, s^{-1}}$. 
Clearly, this estimate does not consider the presence of a top wall and how it reduces the particle angular rotation, as we will see later.

Now let's consider the general situation of a rotating free particle, far away from any surface. The rotational motion is reciprocal, and unable to induce any drift. However, if close to a solid wall, as shown in the left schematic in Fig.~\ref{fig:figure2}(a), 
the rotational motion can be rectified into a net drift velocity due to the hydrodynamic interaction with the close substrate,~\cite{Goldman1967} which breaks the spatial symmetry of the flow.
The situation changes when the particle is squeezed between two flat channels, as in the middle of Fig.~\ref{fig:figure2}(a). 
Now the asymmetric flow close to the bottom surface is balanced by the one at the top, and the spatial symmetry is restored. As a consequence, the particle is unable to propel unless small thermal fluctuations can 
displace the colloid from the central position making it closer to one of the two surfaces. However, 
even in this case, the rectification will be balanced on average.
Here we show that, in contrast to previous cases, when the particle is confined between two curved walls, the effect of curvature breaks again the spatial symmetry, and induce a net particle motion, as shown in the last schematic in Fig.~\ref{fig:figure2}(a).

\begin{figure*}[ht]
\centering
\includegraphics[width=\textwidth,keepaspectratio]{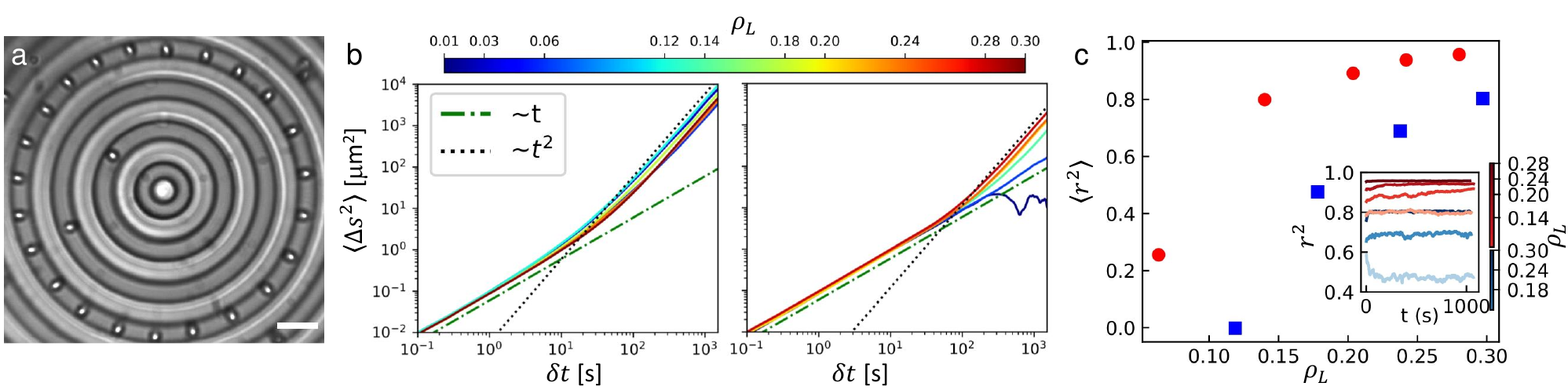}
\caption{(a) Optical microscope image of paramagnetic colloidal particles spinning inside a circular channel with radius $R=35 \, \rm{\mu m}$. The scale bar is 10 $\mu m$, see also 
Movie S2\dag  in the Supporting Information. (b) Mean squared displacement of the circumference arc $\langle \Delta s^2 \rangle$ as a function of the time intervals $\delta t$ for different densities $\rho_L$. Left (right) plot shows the experimental data for a channel with radius  $R = 15 \mu m$ ($R = 35 \mu m$ respectively). The dashed-dotted line indicate the normal diffusion ($\alpha = 1$) while the dotted line shows the ballistic regime ($\alpha = 2$). Colorbar, legend, and axes are common in both plots. (c) Average Kuramoto order parameter $\langle r^2 \rangle$ (Eq.~\eqref{kuramoto}) versus particle density $\rho_L$. Blue squares (red disks) are experimental data for a channel with radius $R = 15 \, \rm{\mu m}$ ($R = 35 \, \rm{\mu m}$ respectively).
Inset shows the temporal evolution of  $\langle r^2 \rangle$ for  different particle densities, identified by the color bar. The color code of the color bar is the same as in the main plot.}
\label{fig:figure4}
\end{figure*}
 
To demonstrate this point, we perform a series of experiments by varying the channel radius $R\in [5,40] \, \rm{\mu m}$ 
and measuring the tangential velocity of the particle $v_{\phi}$, Fig.~\ref{fig:figure2}(b).
We find that while for large rings ($R>20 \, \rm{\mu m}$)
the particle does not display any net movement, $v_{\phi}\sim 0$, below $R\sim 10 \rm{\mu m}$ the effect of curvature becomes important and it can rectify the rotational motion of the particle. Indeed, for $R=5  \rm{\mu m}$ the particle circulates along the channel at a non-negligible speed of $v_{\phi} \simeq 0.14 \rm{\mu m \, s^{-1}}$. 
The rectification process decreases by increasing $R$, while the sense of rotation is equal to that of the precessing field,
sign that the spinning particle is more strongly coupled to the bottom wall of the curved channel. 

This can be also appreciated from the Video S1\dag in the Supporting Information, where a driven particle in the small, $R=5 \, \rm{\mu m}$ ring moves clockwise, even if the velocity is not uniform. Regarding this last point, we note that due to the resolution limit of the lithographic technique,
our small channels may present imperfections such as deformations and non-uniform edges. 
These defects may pin a rotating particle during its motion for a longer time. This effect can be observed in the inset to Fig. \ref{fig:figure2}, where we have 
discretized space and calculated the mean tangential velocity $v_\phi$ for each slice of the circular trajectory. Here the 
time spent by the particle near $\phi=0$ is larger, and at this point, the colloidal particle is trapped by a defect, taking a long time to be released by thermal motion. To avoid this problem, we have calculated the mean velocity directly from the binned series shown in the inset, giving equal weights to all the discrete slices. 

We can use the equation
\begin{align}
v_\phi= M_{v_\phi, \tau_z} \tau_z
\end{align}
to calculate the tangential velocity, where the entry of the mobility matrix, $M_{v_\phi, \tau_z}$, is calculated from finite element simulations. To match the experimental value we fit the value of  $\tau_z$, which is constant for all values of the curvature $R$. 
We have performed a non-linear regression of the experimental data and fit the velocity profile using the torque as a single, free parameter. The calculated off-axis component of the mobility matrix $\bm{M}$ is multiplied by a constant factor, which gives the continuous line shown in Fig.~\ref{fig:figure2}(b). The data shows excellent agreement with the simulation for a torque of $\tau$ = 0.55 pN$\mu$m, which is slightly lower than the calculated value of the magnetic torque,  $\tau_m = 0.73 \, \rm{pN \, \mu m}$. We also performed simulations with the sphere center placed away from the channel centerline and closer to the inner or the outer wall. However, these simulations did not agree well with the experimental data.
The small difference can be attributed to the different approximations used to calculate $\tau_m$, such as the presence of a single relaxation time or the value of the magnetic volume susceptibility of the particle which is a factor that can change from one batch to the other since it depends on the chemical synthesis used to produce these particles. 
 
To understand the origin of the propulsion, we have used the simulation to calculate the surface traction along a sphere rotating at a constant angular velocity. The traction is computed in each point on the particle surface from the stress tensor as $\bm{T}\cdot\bm{n}$, where $\bm{n}$ is the unit vector normal to the surface of the sphere. In Fig. \ref{fig:figure3} we are showing only the $\hat{\bm{\phi}}-$component of the traction vector, which points in the direction of the particle's translational motion. In Figs. \ref{fig:figure3}(b) and (c), where we show the normalized traction force $N_\phi$
calculated in the outer and inner hemispheres of the particle respectively, we find that the coupling of the particle surface to the outer wall pushes the particle back towards the negative $\hat{\bm{\phi}}$-direction, corresponding to a negative angular velocity. In contrast, the interaction between the inner hemisphere of the particle, and the inner wall pushes the particle along the forward direction. The last panel, Fig.~\ref{fig:figure3}(d) shows the sum of both interactions, whose integral yields the total hydrodynamic force acting along the $\hat{\bm{\phi}}$ direction. Even though there is a negative component in the center of the circle, which indicates a higher coupling to the outer wall, it is clear that the strongest contribution comes from the two positive lobes where the inner hemisphere coupling is stronger, and the total force along the $\hat{\bm{\phi}}$ axis is positive.
Thus, our finite element simulations confirm that the rectification process of a single rotating particle within a curved channel results only from the hydrodynamic interactions between the particle and the surfaces.

\section{Driven magnetic single file}
We next explore the collective dynamics of an ensemble of interacting spinning colloids that are confined along the circular channel.
Since the lateral width is $w\sim 2a$ the colloidal particles 
cannot pass each other, thus effectively forming a "single file" system.~\cite{Kag92,Karger1992}.
While the particle dynamics in a "passive" single file, driven only by thermal fluctuations, have been a matter of much interest in the past,~\cite{Wei00,Lutz2004,Lin2005}
more scarce are realizations of "driven" single files~\cite{Koppl2006,Karolis2019}, where the particles   
undergo ballistic transport due to external force and collective interactions.~\cite{Taloni2006,Barkai2009,Illien2013,Dolai2020}
In our case, a typical situation is shown in Fig.~\ref{fig:figure4}(a), where $N=23$ particles move within a channel of radius $R=35 \rm{\mu m}$ (linear density $\rho_L= Na/(2\pi R)=0.15$), see also 
Video S2\dag  in the Supporting Information.
 The applied precessing magnetic field 
apart from inducing the spinning motion within the particle is also used to reduce the magnetic dipolar interaction between them. 
In particular, in the point dipole approximation, we can consider
that the interaction potential $U_{dd}$ between two equal moments $\bm{m}_{i,j}$ at relative position $r=|\bm{r}_i-\bm{r}_j|$
is given by:  $U_{dd}=\mu_0/4\pi \{ (\bm{m}_i \cdot \bm{m}_j/r^3)-[3(\bm{m}_i \cdot \bm{r}) (\bm{m}_j \cdot \bm{r})/r^5]   \} $.
Here the induced moment is given by, $m=V\chi H_0$, being $V=(4\pi a^3)/3$ the particle volume. If we assume that the induced moments follow 
the precessing field with a cone angle $\theta$, 
these interactions can be time-averaged, and become:
\begin{equation}
\langle U_{dd} \rangle =\frac{\mu_0 m^2}{4\pi r^3}P_2(\cos{\theta})\, \, \, ,
\label{dipolarint}
\end{equation}
where $P_2(\cos{\theta})=(3\cos^2{\theta}-1)/2$ is the second order Legendre polynomial. 
For the field parameters used, the dipolar interaction is small since $\theta=44.5^{\circ}$ which is close to the "magic angle" ($\theta=45.7^{\circ}$)
where $P_2(\cos{\theta})\sim 0$. From these values, we estimate that the pair repulsion 
is of the order of $U_{dd}= 5 k_B T $ 
being $T= 293 \, \rm{K}$ the thermodynamic temperature and 
$k_B$ the Boltzmann constant. Thus, the magnetic interactions are slightly repulsive to avoid the particles
attracting each other at high density. 
Indeed, for a purely rotating magnetic field ($H_z=0$, $\theta = 90^{\circ}$) the spinning particles  
aggregate forming compact chains that induce clogging of the channel. 
In contrast, to obtain a strong repulsion we would require a smaller in-plane field component that would induce a smaller magnetic torque.
  
Thus, we perform different experiments by varying both the 
particle linear density within the ring, $\rho_L$, and the channel radius $R$, while keeping fixed the applied field parameters, which are the same as in the single-particle case. 
From the particle positions, we measure the mean squared displacement (MSD) along the circumference arc $\Delta s$ connecting a pair of particles within the ring. 
Here the MSD is calculated as $\langle \Delta s^2 (\Delta t) \rangle \equiv \langle (s(t)-s(t+\delta t))^2 \rangle $, being $\delta t$ the lag time and $\langle … \rangle$ a time average. 
We use the MSD to distinguish the dynamic regimes at the steady state. Here the MSD behaves as a power law $\langle \Delta s^2 (\Delta t) \rangle \sim \tau^{\alpha}$, with an exponent $\alpha=1 $ for standard diffusion and $\alpha = 2$ for ballistic transport, i.e. a single file of particles with constant speed.
As shown in the left plot of Fig.~\ref{fig:figure4}(b)  
for a small channel width, $R=15 \, \rm{\mu m}$,
all MSDs behave similarly, showing an initial diffusive regime followed by a ballistic one at a longer time, and 
the transition between both dynamic regimes occurs at $\sim 10-20$s.
Increasing the particle density has a small effect in the ballistic regime, 
where the velocity is higher at a lower density. This effect indicates that  the inter-particle interactions
reduce the average speed of the single file. 
Since magnetic dipolar interactions are minimized by spinning the field at the "magic angle", this effect may arise from hydrodynamics. Indeed it was recently observed that these interactions may affect the circular motion of driven systems.~\cite{Cereceda2021,Lips2022} 
In contrast, the situation changes for a larger channel, $R=35 \, \rm{\mu m}$ and right plot in Fig.~\ref{fig:figure4}(b).
There, at low density $\rho_L$ the particle tangential speed vanishes, $v_{\phi}=0$ (as shown in Fig.~\ref{fig:figure2}(b)), and thus we recover a diffusive regime also at long time until the slope fluctuates due to lack of statistical averages. 
Increasing $\rho_L$ we recover the transition from the
diffusive to the ballistic one, but occurring at a relatively longer time, $\sim 100$s.
The emergence of a ballistic regime results from the cooperative movement of the particles which generate hydrodynamic flow fields near the surface. Indeed, while single particles display negligible translational velocity in channel of radii $R= 15\rm{\mu m}$ and $35 \rm{\mu m}$ (Fig. 2(a)), we observed a ballistic regime in both cases at high density. As previously observed for a linear, one-dimensional chain of rotors~\cite{martinez-pedrero_colloidal_2015}, the spinning particles generate a cooperative flow close to a bounding wall which act as a “conveyor belt effect”. Here, the situation is similar, however the larger distance between the particles due to their repulsive interactions and the presence of two walls which screen the flow reduce the average speed of the single file and increase the time required for this cooperative effect to set in.

We can estimate the transition time between both regimes
for the single particle. We start by 
measuring the diffusion coefficient of the single particle within the channel from the MSD in the diffusive regime, $D= 0.0379 \pm 0.0002 \rm{\mu m^2 \, s^{-1}}$.
Thus, the motion of an individual magnetic particle will turn from the Brownian motion with $\langle \Delta s^2  \rangle = 2 D t$  to a ballistic one 
with $\langle \Delta s^2  \rangle = (R \omega_p t)^2$
after a characteristic time $\tau_c=D/(R\omega_p)^2$.
From the experimental data in Fig.~\ref{fig:figure2}(b) we have that 
$\omega_p=0.0073 \, \rm{rad \, s^{-1}}$ for $R= 15 \, \rm{\mu m}$ which gives $\tau_c=3.1$ s, similar to the small channel in the left of Fig.~\ref{fig:figure4}(b).

We further analyze the relative particle displacement, 
and the presence of synchronized movement by
measuring a Kuramoto-like~\cite{Kuramoto1987,Strogatz2000} order parameter,
defined as:
\begin{equation}
\langle r^2(t)\rangle  = \frac{1}{N^2} \sum_{i, j = 1}^N \cos{ ( \Delta \varphi_{ij} (t) )} 
\label{kuramoto}
\end{equation}
where $N$ is the number of particles, $\Delta \varphi_{ij}=\varphi_i - \varphi_j$ the difference between the angular position of two nearest particles $i,j$ and $\varphi_{i,j}$
is measured with respect to one axis ($\hat{\bm{x}}$) located in the particle plane. This order parameter is
normalized such that $\langle r^2 \rangle \in [0,1]$, where $1$
corresponds to the fully coordinated relative movement of all spinning particles.
Fig.~\ref{fig:figure4}(c) show the average value
of $r^2$ at different density $\rho_L$. In particular, we find that 
$r^2$ increases with $\rho_L$ since,
for a large number of particles the repulsive particles start to interact with each other and their translation becomes more coordinated.
On the other hand, we find that $\langle r^2 \rangle$ decreases with the  
radius of curvature (data from red from blue)
meaning that the curvature reduces the cooperative translation between the spinning particles. 
The cooperative movement revealed from the Kuramoto order parameter results from the repulsive magnetic dipolar interactions. Indeed, this parameter increases with density i.e. when the particles are closer to each other and experience stronger repulsive forces. However, the emergence of self-propulsion in our system is a pure hydrodynamic effect induced by the wall curvature. Indeed, the precessing field induces reciprocal magnetic dipolar interactions which could not give rise to any net motion of the spinning particles.

\section*{Conclusions}
We have demonstrated that a circular channel can be used to induce and enhance the propulsion of spinning particles while the effect becomes absent for a flat one. The propulsion appears from the different hydrodynamic coupling to each of the two curved walls; even if the no-slip condition of each wall pushes the particle along an opposite direction, the two interactions are not symmetric due to the channel curvature. The dominating interaction is the one from the inner wall, which produces a rotation along the channel along the same direction as the angular velocity. We confirm this hypothesis both via experiments and finite element simulations. 

We note here that the realization of field-driven magnetic prototypes able to propel in viscous fluids is
important for both fundamental and technological reasons. 
In the first case, the collection of self-propelled can be used as 
a model system to investigate fundamental aspects of 
non-equilibrium statistical physics.~\cite{Ramaswamy2010,Marchetti2013,Fodor2016,Nardini2017,palacios_guided_2021} On the other hand, 
it is expected that these prototypes can be used as controllable inclusions in microfluidic systems,~\cite{Tierno2008,Burdick2008,Shakuntala2008,Sanchez2011,Sharan2021} either to control or regulate  
the flow~\cite{Terray2002,Sawetzki2008,Kavcic2009} or to transport microscopic 
biological or chemical cargoes.~\cite{Zhang2010,Palacci2013,Gao2014,Martinez-Pedrero2015,Pedrero2017,Junot2023,Li2023} 
The propulsion mechanism unveiled in this work could be used to move particles along a microfluidic device, either by designing it with a circular path, or by coupling many semicircles in a zig-zag to form a straight path. 

%

\section*{Data availability}
The data that support the findings of this study are available
from the corresponding authors upon reasonable request.

\section*{Conflicts of interest}
There are no conflicts to declare.

\section*{Acknowledgements}
We thank Carolina Rodriguez-Gallo who shared the protocol for the sample preparation, and the MicroFabSpace and Microscopy Characterization Facility, Unit 7 of ICTS “NANBIOSIS” from CIBER-BBN at IBEC. 
This project has received funding from the European Research Council (ERC) under the European Union's Horizon 2020 research and innovation program (grant agreement no. 811234).
M.D.C. acknowledges funding from the Ramon y Cajal fellowship RYC2021-030948-I and the grant PID2022-139803NB-I00 funded by the MICIU/AEI/10.13039/501100011033 and by the EU under the NextGeneration EU/PRTR \& FEDER 
programs.
I. P. acknowledges support from  MICINN and DURSI for financial support under Projects No. PID2021-126570NB-100 AEI/FEDER-EU and No. 2021SGR-673, respectively, and from Generalitat de Catalunya (ICREA Acad\'emia).
P. T. acknowledges support from the 
Ministerio de Ciencia, Innovaci\'on y Universidades (grant no. PID2022-
137713NB-C21 AEI/FEDER-EU) and the Ag\`encia de Gesti\'o d'Ajuts Universitaris
i de Recerca (project 2021 SGR 00450) and the Generalitat de
Catalunya (ICREA Acad\'emia).



\balance


\bibliography{RotorsInChannels}
\bibliographystyle{rsc} 

\end{document}